\documentclass[aps,prd,onecolumn,groupedaddress,floatfix,longbibliography,superscriptaddress,notitlepage]{revtex4-1}

\usepackage{amsmath}
\usepackage{amssymb}
\usepackage{amsfonts}
\usepackage{graphicx}
\usepackage{bbold}
\usepackage{bm}
\usepackage{bm}
\usepackage{cancel}
\usepackage{times,float}
\usepackage{graphicx}
\usepackage[usenames,dvipsnames,svgnames]{xcolor}
\usepackage{hyperref}
\hypersetup{colorlinks=true, linkcolor=NavyBlue, citecolor=PineGreen,urlcolor=cyan}
\usepackage{multirow}
\usepackage{physics}
\usepackage{esint}
\usepackage{caption}
\usepackage{subcaption}
\usepackage{tikz}
\usepackage{pgfplots}
\usepackage{pgfplotstable}
\pgfplotsset{compat = newest}
\usetikzlibrary{positioning, arrows.meta}
\usepgfplotslibrary{fillbetween}
\usepackage{svg}

\begin{document}

\title{Casimir effect between semitransparent mirrors in a Lorentz-violating background}

\author{Rom\'an Linares}
\email{lirr@xanum.uam.mx}
\affiliation{Departamento  de  F\'isica,  Universidad  Aut\'onoma  Metropolitana-Iztapalapa, San Rafael Atlixco 186, 09340 Ciudad de M\'{e}xico, M\'{e}xico}

\author{C. A. Escobar}
\email{carlos.escobar@xanum.uam.mx}
\affiliation{Departamento  de  F\'isica,  Universidad  Aut\'onoma  Metropolitana-Iztapalapa, San Rafael Atlixco 186, 09340 Ciudad de M\'{e}xico, M\'{e}xico}

\author{A. Mart\'{i}n-Ruiz}
\email{alberto.martin@nucleares.unam.mx}
\affiliation{Instituto de Ciencias Nucleares, Universidad Nacional Aut\'{o}noma de M\'{e}xico, 04510 Ciudad de M\'{e}xico, M\'{e}xico}

\author{E. Pl\'acido}
\email{edyplax2@hotmail.com}
\affiliation{Departamento  de  F\'isica,  Universidad  Aut\'onoma  Metropolitana-Iztapalapa, San Rafael Atlixco 186, 09340 Ciudad de M\'{e}xico, M\'{e}xico}

\begin{abstract}
We investigate the Casimir effect for a massless scalar field confined between two parallel semitransparent mirrors in a vacuum modified by spontaneous Lorentz symmetry breaking. Using Green's function techniques and a point-splitting evaluation of the stress-energy tensor, we compute the vacuum expectation value of the energy density $T _{00}$. After a suitable renormalization, the Casimir energy is obtained as the difference between the vacuum configurations with and without the mirrors. We derive closed-form expressions that generalize the conventional result by simultaneously incorporating the mirror transparency and the Lorentz-violating background. Our analysis shows that the effect of transparency and Lorentz violation consists of a multiplicative correction and an effective rescaling of the plate separation, thereby modifying the functional dependence of the energy on the distance. Beyond the formal derivation, we discuss possible physical realizations of this framework, emphasizing anisotropic media such as nematic liquid crystals (e.g., 5CB), where uniaxial dielectric properties could emulate the Lorentz-violating background. Numerical estimates for such systems illustrate the phenomenological impact of our results and open the possibility of constraining Lorentz-violating coefficients through precision Casimir measurements.
\end{abstract}

\maketitle

\section{Introduction} \label{Intro}

The Casimir effect is one of the most striking macroscopic manifestations of quantum field theory, arising from zero-point vacuum fluctuations in the presence of material boundaries \cite{Casimir1948}. First predicted by Casimir in 1948 for two perfectly conducting parallel plates, and subsequently measured by Sparnaay a decade later \cite{Sparnaay1958}, it has since become a benchmark phenomenon in both theoretical and experimental physics. Over the years, the effect has been generalized to a wide range of field contents, geometries \cite{PhysRevLett.99.170403,PhysRevLett.88.101801,PhysRevE.80.021125,Capasso}, and boundary conditions \cite{Juárez-Aubry_2021, Juárez-Aubry_Weder_2022}, and continues to serve as a sensitive probe for new physics beyond the Standard Model \cite{Klimchitskaya2023, PhysRevD.100.116023,PhysRevD.101.046023}.

Among the theoretical frameworks proposed to incorporate possible departures from Lorentz invariance, the Standard-Model Extension (SME) stands out as the most systematic and comprehensive. In this formulation, deviations from exact Lorentz symmetry are encoded in constant background tensors, which act as effective parameters capturing the strength and directionality of the violation \cite{PhysRevD.55.6760, PhysRevD.58.116002}. These coefficients propagate their influence across all sectors of the theory, allowing one to trace their imprints in a broad range of phenomena. Within this context, the scalar sector \cite{EDWARDS2018319} has proven to be particularly instructive for exploring Casimir-type effects, where modifications to the vacuum stress and energy density can be obtained in closed form \cite{PhysRevD.101.095011, ESCOBAR2020135567, PhysRevD.102.015027, doi:10.1142/S0217751X21501682}. Beyond the scalar case, Lorentz-violating corrections have also been investigated in fermionic \cite{PhysRevD.99.085012, 10.1093/ptep/ptae016}, electromagnetic \cite{PhysRevD.94.076010, PhysRevD.95.036011}, and gravitational settings \cite{Sorge_2005, PhysRevD.89.024015, Nazari_EPJC_2015}, as well as in scenarios involving nontrivial geometries, finite temperature, and higher-derivative interactions. Collectively, these studies highlight the Casimir effect as a versatile arena in which Lorentz-violating physics can be both modeled and constrained.

Despite this progress, most Lorentz-violating analyses of the Casimir effect have assumed perfectly reflecting boundaries, thereby neglecting the important role of mirror transparency. Yet, recent studies on light propagation and related phenomena in the presence of semitransparent mirrors have demonstrated that transparency parameters can strongly affect the sensitivity to Lorentz-violating coefficients \cite{Escobar:2023huu}. This opens the possibility of exploring new experimental windows where Lorentz violation could leave observable imprints.

In this work, we address this gap by investigating the Casimir effect for a real scalar field in the presence of both semitransparent mirrors and explicit Lorentz-violating corrections in the scalar sector of the SME. Each mirror is modeled by a delta-function potential characterized by its strength, which interpolates between full transparency and perfect reflection. Lorentz violation is encoded in a constant, symmetric background tensor $h ^{\mu\nu}$, which modifies the effective dispersion relations of the field. By employing Green's function methods and a suitable renormalization scheme, we derive closed-form expressions for the Casimir energy density $T _{00}$, incorporating the combined influence of mirror transparency and Lorentz-violating effects. The resulting formulae reduce smoothly to the conventional Casimir energy in the Lorentz-invariant, perfect-mirror limit, while revealing multiplicative corrections that can be traced independently to each parameter.

Finally, we complement our formal analysis with realistic physical estimates, connecting the Lorentz-violating background tensor to effective optical metrics in anisotropic condensed-matter systems. In particular, we show that nematic liquid crystals such as 5CB (pentyl-cyanobiphenyl) provide a feasible scenario where anisotropies of the refractive index naturally generate $h ^{\mu\nu}$ coefficients of significant magnitude, offering a concrete experimental platform to realize our theoretical framework. This dual perspective, from fundamental field-theoretic motivation to condensed-matter realizations, highlights the potential of Casimir experiments as a bridge between high-energy phenomenology and laboratory-scale physics.

The remainder of this paper is organized as follows. In Sec. \ref{Model_Method_Section} we introduce the model, define the semitransparent boundary conditions, and outline the point-splitting procedure used to evaluate the stress-energy tensor. Section \ref{Lorentz_invariant_Section} revisits the Casimir effect in the Lorentz-invariant setting, establishing the baseline results for partially transmitting mirrors. In Section \ref{Lorentz_violation_Section} we extend the analysis to a Lorentz-violating background, deriving the corresponding modifications to the Casimir energy. Section \ref{Physical_realizations} is devoted to possible physical realizations, with an emphasis on anisotropic media such as nematic liquid crystals, and provides numerical illustrations of the parameter dependence of our results. Finally, Section \ref{Conclusions} summarizes our findings and discusses their relevance for future experimental constraints on Lorentz-violating coefficients. Throughout the paper, we use the Minkowski metric with signature $\eta ^{\mu \nu} = \textrm{diag} (+,-,-,-)$. In what follows, we set $\hbar = c = 1$, restoring them when presenting the final expressions for the Casimir energy density.

\section{Model definition and point-splitting evaluation of the stress-energy tensor} \label{Model_Method_Section}

The Lorentz symmetry violation considered in this work originates from the scalar sector of the Standard-Model Extension (SME) for a massless real scalar field $\phi (x ^{\mu})$  \cite{EDWARDS2018319}. In addition to the Lorentz-violating terms, the model includes delta-function contributions that represent the interaction of the scalar field with two parallel semitransparent mirrors. These localized terms effectively model the mirrors as partially reflective surfaces and allow for a consistent treatment of boundary conditions within the field-theoretic formalism. The complete Lagrangian density describing this setup is given by
\begin{align}
    \mathcal{L} = \frac{1}{2} h ^{\mu \nu} \, ( \partial _{\mu} \phi ) \, ( \partial _{\nu} \phi ) - \frac{1}{2} \lambda \,  \delta (z) \, \phi ^{2} - \frac{1}{2} \lambda ^{\prime} \, \delta (z-L) \, \phi ^{2} , \label{Lagangian}
\end{align}
where $h ^{\mu\nu}$ is a constant background field that encodes the effects of Lorentz symmetry breaking. It arises from a nontrivial vacuum structure and, importantly, does not transform as a rank-two tensor under active Lorentz transformations, reflecting the presence of a preferred spacetime direction in the underlying theory. The delta-function terms $\delta (z)$ and $\delta (z-L)$ represent two parallel semitransparent mirrors located at $z = 0$ and $z = L$, respectively, with coupling constants $\lambda$ and $\lambda '$ that quantify the degree of transparency of each mirror. In the limit $\lambda, \lambda ' \to \infty$, the mirrors become perfectly reflecting, recovering the conventional Dirichlet boundary conditions. This Lagrangian allows us to explore how both the Lorentz-violating background and the partial transparency of the mirrors affect the structure of the quantum vacuum and, in particular, the Casimir energy.

It is well established in the SME literature that the term $\frac{1}{2} h ^{\mu \nu} \, ( \partial _{\mu} \phi ) \, ( \partial _{\nu} \phi )$ can be mapped onto the standard kinetic term for a scalar field via suitable field and coordinate redefinitions. This mapping is typically valid to first order in the expansion $h^{\mu\nu}=\eta^{\mu\nu}+\delta h ^{\mu\nu}$, which is sufficient for most phenomenological studies, given the tight experimental constraints on Lorentz-violating coefficients. Nonetheless, by adopting a fully non-perturbative treatment, our analysis remains valid beyond the linear regime, making the results applicable not only to theories with explicit Lorentz violation but also to a broader class of models involving anisotropic or direction-dependent modifications to field dynamics.

The vacuum expectation value (VEV) of the stress-energy tensor, $T ^{\mu \nu} = h ^{\mu \alpha} \partial _{\alpha} \phi \partial ^{\nu} \phi - \eta ^{\mu \nu} \mathcal{L}$,
plays a central role in the computation of the Casimir energy. This quantity can be evaluated using Green's function (GF) techniques, which provide an efficient and systematic approach to compute vacuum fluctuations in the presence of nontrivial backgrounds and boundary conditions \cite{doi:10.1142/4505}. To illustrate the method, we begin by recalling that the Green's function is defined as the vacuum expectation value of the time-ordered product of field operators, i.e., the two-point correlation function \cite{PhysRevD.20.3063, Peskin:1995ev},
\begin{align}
    G(x,x') = - i \, \langle 0 \! \mid \! \mathcal{T} \left\lbrace \phi(x) \phi (x') \right\rbrace \! \mid \! 0 \rangle  , \label{Green_two_point}
\end{align}
where $\mathcal{T}$ denotes the time-ordering operator, which arranges the field operators such that the one evaluated at the later time appears to the left. From the equations of motion of the Lagrangian in Eq. (\ref{Lagangian}), we find that the Green's function (\ref{Green_two_point}) satisfies the differential equation
\begin{align}
  \left[ h^{\mu\nu}\partial_\mu\partial_\nu+\lambda \delta(z)+\lambda^\prime\delta(z-L) \right] G(x,x') = \delta (x-x') . \label{Green_equation}
\end{align}
This function captures the response of the quantum vacuum to localized sources and encodes all information needed to compute the VEV of composite operators such as $T ^{\mu \nu}$. The point-splitting method can then be employed to express $\langle T^{\mu\nu}\rangle$ in terms of derivatives of $G(x,x')$, and to isolate and subtract divergences in the coincidence limit $x \to x'$, ultimately yielding the renormalized Casimir energy. The VEV of the stress-energy tensor then takes the form \cite{PhysRevD.20.3063, Peskin:1995ev}
\begin{align}
    \langle T ^{\mu \nu} \rangle = - i \lim _{x \rightarrow x'}   h ^{\mu \alpha} \partial _{\alpha} \partial ^{\prime \nu}     G(x,x^\prime) - \eta ^{\mu \nu} \, \langle \mathcal{L} \rangle  , \label{VEV_energy_momentum}
\end{align}
where
\begin{align}
    \langle \mathcal{L} \rangle = \frac{i}{2}  \lim _{x \rightarrow x'}    h ^{\alpha \beta} \, \partial _{\alpha} \partial ' _{\beta}   G(x,x^\prime) . \label{VEV_Lagrangian}
\end{align}
Here, the limit is taken after differentiation.

To proceed with our analysis, we now take a step further and introduce the physical configuration under consideration. Specifically, we study the Casimir effect in the standard setup of two parallel plates separated by a distance $L$, which we choose to be aligned perpendicularly to the $z$-axis. This classical geometry allows us to implement a dimensional reduction by exploiting the translational symmetry along the directions parallel to the plates. The translational invariance in the $(t,x,y)$ directions enables a (2+1)-dimensional Fourier decomposition of the Green's function, given by \cite{doi:10.1142/4505}
\begin{align}
    G(x,x') = \int \frac{d \omega}{2 \pi } \int \frac{d ^{2} \mathbf{k} _{\perp} }{( 2 \pi ) ^{2} } \, e ^{i \mathbf{k} _{\perp} \cdot ( \mathbf{x} _{\perp} - \mathbf{x}' _{\perp} ) }  e ^{- i \omega (t-t') } \, g \left( z , z' ; \omega , \mathbf{k} _{\perp} \right) , \label{Green_function}
\end{align}
where $\mathbf{k} _{\perp} = (k_{x},k_{y})$ and $\mathbf{x} _{\perp} =(x,y)$ denote the momentum and position components in the directions parallel to the plates, respectively. The reduced Green's function $g \left( z , z' ; \omega , \mathbf{k} _{\perp} \right) $ captures the nontrivial behavior in the $z$-direction, where translational invariance is broken due to the presence of the plates. This decomposition simplifies the problem significantly and allows us to isolate the effects of boundary conditions and Lorentz violation along the confinement axis.

Since the configuration is translationally invariant in the $t,x,y$ directions, we take the coincidence limit $t=t'$ and $\mathbf{x} _{\perp} = \mathbf{x}' _{\perp}$ when evaluating the vacuum expectation value of the stress-energy tensor, so that the exponential factors drop out. We then define the effective differential operators acting on the reduced Green's function $g \left( z , z' ; \omega , \mathbf{k} _{\perp} \right) $ as
\begin{align}
    \hat{\partial} ^{\mu} = (-i \omega , - i k ^{1} , - i k ^{2} , - \partial _{z}) , \qquad \hat{\partial} ^{\prime \mu} = (i \omega , i k ^{1} , i k ^{2} , -\partial _{z'}). 
\end{align}
These expressions account for the action of spacetime derivatives on the Fourier representation of $G(x,x')$. As a result, the VEV of the stress-energy tensor becomes
\begin{align}
    \langle T ^{\mu \nu} \rangle = - i \lim _{z \rightarrow z'} \int \frac{d \omega}{2 \pi } \int \frac{d ^{2} \mathbf{k} _{\perp} }{( 2 \pi ) ^{2} } \,    h ^{\mu \alpha} \hat{\partial} _{\alpha} \hat{\partial} ^{\prime \nu}  \,  g \left( z , z' ; \omega , \mathbf{k} _{\perp} \right) - \eta ^{\mu \nu} \, \langle \mathcal{L} \rangle  , \label{VEV_energy_momentum_reduced}
\end{align}
where
\begin{align}
    \langle \mathcal{L} \rangle = \frac{i}{2}   \lim _{z \rightarrow z'} \int \frac{d \omega}{2 \pi } \int \frac{d ^{2} \mathbf{k} _{\perp} }{( 2 \pi ) ^{2} } \,     h ^{\alpha \beta} \, \hat{\partial} _{\alpha} \hat{\partial} ' _{\beta} \,  g \left( z , z' ; \omega , \mathbf{k} _{\perp} \right) . 
\end{align}
This expression isolates the relevant dynamics along the $z$-direction, where the boundary conditions of the Casimir configuration are imposed. The integration over $\omega$ and $\mathbf{k} _{\perp}$ will be carried out once the reduced Green's function $g \left( z , z' ; \omega , \mathbf{k} _{\perp} \right)$ is determined. To this end, we insert the Fourier representation of the full Green's function (\ref{Green_function}) into the defining equation (\ref{Green_equation}) and exploit the translational invariance along the $t$, $x$ and $y$ directions to isolate the dependence on the $z$-coordinate. This yields a differential equation for the reduced Green's function of the form
\begin{align}
  \left[ h ^{\mu \nu} \hat{\partial} _{\mu} \hat{\partial} _{\nu} + \lambda \delta(z) + \lambda' \delta(z-L) \right] g \left( z , z' ; \omega , \mathbf{k} _{\perp} \right)  = \delta (z-z') . \label{Reduced_Green_equation}
\end{align}
In the following sections, we apply the formalism introduced above to study the Casimir effect in the presence of semitransparent mirrors. First, in the Lorentz-invariant case, which serves as a baseline, and then in Section \ref{Lorentz_invariant_Section}, we explore how Lorentz-symmetry violation modifies the vacuum energy.

\section{Lorentz-invariant Casimir effect with semitransparent boundaries} \label{Lorentz_invariant_Section}

We begin by analyzing the Casimir effect in a Lorentz-invariant setting (i.e. we take $h ^{\mu \nu} \to \eta ^{\mu \nu}$), where a real scalar field is confined between two parallel semitransparent mirrors. This case provides a reference for assessing Lorentz-violating modifications discussed in the following section. To proceed, we construct the reduced Green’s function $g \left( z , z' \right)  \equiv g \left( z , z' ; \omega , \mathbf{k} _{\perp} \right)  $, defined as the solution to the differential equation governing the system in the presence of two semitransparent mirrors located at $z=0$ and $z=L$. The equation satisfied by $g$ reads
\begin{align}
  \left[ - \omega ^{2} + k _{\perp} ^{2} - \partial _{z} ^{2}  + \lambda \, \delta(z) + \lambda' \, \delta(z-L) \right] g \left( z , z' \right)  = \delta (z-z') ,\label{Reduced_Green_equation_Lor_Inv}
\end{align}
where $k _{\perp} ^{2} = \mathbf{k} _{\perp} \cdot \mathbf{k} _{\perp} $. This function must satisfy the continuity and jump conditions derived from the properties of the boundaries. 

To solve the reduced Green's function equation (\ref{Reduced_Green_equation_Lor_Inv}) we adopt a method analogous to that used in quantum mechanics for solving the Schrödinger equation in the presence of point interactions. Specifically, we exploit the properties of the Dirac delta function and express the full Green's function $g \left( z , z' \right)$ in terms of the free (mirrorless) Green's function $g _{0} \left( z , z' \right)$, which satisfies
\begin{align}
  \left( - \omega ^{2} + k _{\perp} ^{2} - \partial _{z} ^{2}  \right) g _{0} \left( z , z' \right)  = \delta (z-z') .  \label{Reduced_Green_equation_Lor_Inv_Mirrorless}
\end{align}
The free Green's function $g _{0} (z,z')$, which solves the homogeneous equation in the absence of the delta-function interactions, is well known and takes the form:
\begin{align}
    g _{0} (z,z') = \frac{i}{2 \gamma} e ^{i\gamma (z _{>} -z _{<} ) } , \label{Green_function_0th_order}
\end{align}
where $z_{<} = \mbox{min} (z,z')$, $z_{>} = \mbox{max} (z,z')$ and we define $\gamma = \sqrt{\omega ^{2} - k _{\perp} ^{2} + i \epsilon }$, with the standard $i \epsilon$ prescription ensuring causal (Feynman) boundary conditions. 

The delta-function potentials in Eq. (\ref{Reduced_Green_equation_Lor_Inv_Mirrorless}) allow us to derive simple matching conditions across the interaction points, which will be used to construct the full solution. This approach not only simplifies the calculation but also makes the physical interpretation of the boundary interactions more transparent. By directly integrating the full Green's equation (\ref{Reduced_Green_equation_Lor_Inv}) one obtains an exact expression for the complete Green's function $g \left( z , z' \right)$ in terms of the free Green's function $g _{0} \left( z , z' \right)$:
\begin{align}
    g(z,z') = g _{0} (z,z') - \lambda \, g(z,0) g _{0} (0,z') - \lambda ' \, g(z,L) g _{0} (L,z') . \label{Reduced_Green_1rst_Integral}
\end{align}
To fully determine the reduced Green's function $g(z,z')$, we evaluate Eq. (\ref{Reduced_Green_1rst_Integral}) at the two boundary positions $z=0$ and $z=L$, obtaining two coupled algebraic equations for $g(0,z')$ and $g(L,z')$. These can then be solved explicitly and substituted back into the general expression for $g(z,z')$. The final result is
\begin{align}
    g (z,z') = g _{0} (z,z') & - \frac{1}{\Delta} \lambda \, \{[1 + \lambda ' \, g _{0}(0,0)] \, g _{0}(z,0)  - \lambda ' \, g _{0} (0,L) \, g _{0}(z,L) \} \,  g _{0} (0,z') \notag \\[4pt] & - \frac{1}{\Delta} \lambda ' \, \{ - \lambda \, g _{0} (0,L) \, g _{0} (z,0) + [ 1 + \lambda \, g _{0} (0,0) ] \, g _{0} (z,L) \} \, g _{0} (L,z') , \label{Reduced_Green_final}
\end{align}
where $\Delta = [ 1 + \lambda \, g _{0} (0,0)] \, [1 + \lambda ' \, g _{0} (0,0) ] - \lambda \, \lambda ' \, g _{0} (0,L) \, g _{0} (0,L)$. To derive this result we used the properties  $g _{0} (0,0) = g _{0} (L,L)$ and $g _{0} (0,L) = g _{0} (L,0)$.

To obtain the physically meaningful (renormalized) vacuum energy, we subtract the zero-point energy of the free theory from that of the bounded configuration \cite{PhysRevD.20.3063, PhysRev.184.1272}. This procedure isolates the finite contribution induced by the presence of the plates. In the quantum field theory framework, the renormalized Casimir energy per unit area is then computed in terms of the VEV of the stress-energy tensor as
\begin{align}
    E _{C} (L) = \int ^{L} _{0} \left( \langle T ^{00} \rangle _{\parallel} - \langle T ^{00} \rangle _{v} \right) dz , \label{Casimir_energy}
\end{align}
where $\langle T ^{00} \rangle _{\parallel}$ denotes the vacuum energy density in the presence of the boundaries and $\langle T ^{00} \rangle _{v}$ corresponds to the VEV in the absence of any plates. From Eq. (\ref{VEV_energy_momentum_reduced}) we read the expression for the vacuum energy density as
\begin{align}
    \langle T ^{00} \rangle = \frac{1}{i} \, \lim _{z \rightarrow z'} \, \int \frac{d \omega}{2 \pi } \int \frac{d ^{2} \mathbf{k} _{\perp} }{( 2 \pi ) ^{2} } \,  \omega ^{2} \,  g \left( z , z' \right) - \langle \mathcal{L} \rangle  , \qquad  \langle \mathcal{L} \rangle = \frac{1}{2i} \, \lim _{z \rightarrow z'} \, \int \frac{d \omega}{2 \pi } \int \frac{d ^{2} \mathbf{k} _{\perp} }{( 2 \pi ) ^{2} } \, \left(      \gamma ^{2} -   \partial _{z} \partial ' _{z} \right)  g \left( z , z' \right) . \label{VEV_energy_LS}
\end{align}
Using the free-space reduced Green's function defined in Eq. (\ref{Green_function_0th_order}) and performing a Wick rotation $\omega \rightarrow i \zeta$ we obtain
\begin{align}
    \langle T ^{00} \rangle _{v} =  \int \frac{d \zeta }{2 \pi } \int \frac{d ^{2} \mathbf{k} _{\perp} }{( 2 \pi ) ^{2} } \,  \frac{\zeta ^{2}}{ 2 \tilde{\gamma}  }    , \label{VEV_T00_Lorentz_symmetric}
\end{align}
where $\tilde{\gamma} = \sqrt{ \zeta ^{2} + k _{\perp} ^{2} } $. It is important to emphasize that the VEV of the Lagrangian vanishes, i.e. $\langle \mathcal{L} \rangle =0$. For the sake of clarity, we omit the details of the proof in this section; however, they will be presented later as a limiting case of the Lorentz-violating Lagrangian VEV.

Following a similar procedure, we evaluate the VEV of the energy density in the presence of the semitransparent plates using Eq. (\ref{Reduced_Green_final}). Using the following results (for $0<z,z'<L$)
\begin{align}
    & \lim _{z \rightarrow z'} \, \partial _{z} \partial ' _{z} \, g _{0}(z,0) \,  g _{0} (0,z') = - \gamma ^{2} g _{0}(z,0) \,  g _{0} (0,z)  , \qquad \lim _{z \rightarrow z'} \, \partial _{z} \partial ' _{z} \, g _{0}(z,L) \,  g _{0} (0,z') = \gamma ^{2}  g _{0}(z,L) \,  g _{0} (0,z) , \notag \\ & \lim _{z \rightarrow z'} \, \partial _{z} \partial ' _{z} \, g _{0}(z,L) \,  g _{0} (L,z') = - \gamma ^{2}  g _{0}(z,L) \,  g _{0} (L,z) ,   
\end{align}
we obtain the following expression:

\begin{align}
    \langle T ^{00} \rangle _{\parallel} = \langle T ^{00} \rangle _{v} \, + \frac{1}{2} \,  \int \frac{d \zeta}{2 \pi } \int \frac{d ^{2} \mathbf{k} _{\perp} }{( 2 \pi ) ^{2} } \, \Bigg\{\ \!\! - \frac{ \lambda   \lambda '  \zeta ^{2}  }{2 \Delta \tilde{\gamma} ^{3} }  e ^{- 2 \tilde{\gamma}  L} - \frac{   k _{\perp} ^{2} }{2 \Delta \tilde{\gamma} ^{2} } \left[   \lambda   \left( 1 + \frac{  \lambda ' }{2 \tilde{\gamma} } \right)  e ^{- 2 \tilde{\gamma} z } + \lambda ' \left( 1 + \frac{  \lambda }{2 \tilde{\gamma} } \right)    e ^{- 2 \tilde{\gamma} (L-z) }  \right] \Bigg\}\ , 
\end{align}
where $\Delta = \left( 1 + \frac{\lambda}{2 \tilde{\gamma} }   \right) \, \left( 1 +  \frac{\lambda ' }{2 \tilde{\gamma} } \right) -  \frac{\lambda \lambda ' }{4 \tilde{\gamma} ^{2} }    e ^{- 2 \tilde{\gamma} L }$. With the VEV in hand, we now proceed to calculate the total vacuum energy stored between the plates. Substituting into Eq. (15) and integrating over $z$, while omitting any $L$ independent contributions, yields the following:
\begin{align}
    E _{C} (L) = \frac{1}{2} \int \frac{d \zeta}{2 \pi } \int \frac{d ^{2} \mathbf{k} _{\perp} }{( 2 \pi ) ^{2} } \, \Bigg\{\ \!\! - \frac{ \lambda   \lambda '  \zeta ^{2} L }{2 \Delta \tilde{\gamma} ^{3} }  e ^{- 2 \tilde{\gamma}  L} - \frac{   k _{\perp} ^{2} }{2 \Delta \tilde{\gamma} ^{2} } \left[ \lambda \lambda ' + \tilde{\gamma} (\lambda + \lambda ' ) \right] \frac{1 - e ^{-2 \tilde{\gamma} L} }{2 \tilde{\gamma} ^{2} } \Bigg\}\  . \label{Casimir_energy}
\end{align}
To compute the integral, we begin by expressing the momentum element as $d ^{2} \mathbf{k} _{\perp} = k _{\perp} dk _{\perp} d \theta $ and carry out the angular integration over $\theta$. We then introduce plane polar coordinates in the $(\zeta, k _{\perp})$-plane, defined by $\zeta = \xi \cos \phi$ and $k _{\perp} = \xi \sin \phi$, with $\phi \in [0,\pi ]$ and $\xi \in [0, \infty )$. Finally, we integrate over the angular variable $\phi$. We thus obtain an expression for the Casimir energy in Eq. (\ref{Casimir_energy}) composed of two distinct contributions:

\begin{align}
     E _{C} (L) = \mathcal{E} _{C} (L) + \mathcal{E} _{\infty} (L) , \label{Casimir_energy_fin}
\end{align}
where (recovering $\hbar$ and $c$)
\begin{align}
\mathcal{E} _{C} (L , \Lambda , \Lambda ' ) &=  \frac{  \hbar c  }{ 12 \pi ^{2} L ^{3} } \,  \int _{0} ^{\infty}  d \chi  \,  \chi ^{3}   \left[  1 - \frac{  \left( \Lambda + 2  \chi \right)   \left( \Lambda' + 2 \chi \right)  +  \Lambda \Lambda '  e ^{- 2 \chi }  }{ \left( \Lambda + 2 \chi \right)   \left( \Lambda' + 2 \chi \right)  -  \Lambda \Lambda '   e ^{-2 \chi } }  \right], \label{finite_part}
\end{align}
is finite and encodes the physically meaningful vacuum energy between the plates. The second term,
\begin{align}
\mathcal{E} _{\infty} (L , \Lambda , \Lambda ' ) &= - \frac{ \hbar c  }{ 6 \pi ^{2} L ^{3} } \,  \int _{0} ^{\infty}  d \chi  \,       \frac{  \chi ^{2} \left[   (\Lambda + \Lambda ' ) \chi +  \Lambda \Lambda ' \right] \left( 1 - e ^{-2  \chi }  \right) }{\left( \Lambda + 2  \chi \right)   \left( \Lambda' + 2  \chi \right)  -  \Lambda \Lambda ' e ^{- 2 \chi }}  , \label{divergent_part}
\end{align}
is divergent and it does not contribute to the physically measurable force between the plates. This will be explicitly demonstrated later through the evaluation of the VEV of the normal stress component $T_{zz}$, which determines the Casimir pressure. In these expressions $\Lambda = \lambda L$ and $\Lambda '= \lambda 'L$.

To gain further insight into the physical content of our result, we now examine two important limiting cases: those of perfectly reflecting plates and those of nearly transparent plates. The former corresponds to taking the coupling constants to infinity, $\Lambda, \Lambda' \to \infty$, effectively imposing Dirichlet boundary conditions. The finite vacuum energy becomes
\begin{align}
\lim _{\Lambda, \Lambda' \to \infty } \mathcal{E} _{C} (L , \Lambda , \Lambda' ) &=  \frac{ \hbar c }{ 12 \pi ^{2} L ^{3} } \,  \int _{0} ^{\infty}  d \chi  \,  \chi ^{3}   \left( 1 - \coth \chi \right) = -  \frac{\pi ^{2}  \hbar c  }{1440  L ^{3} }  \label{EC0}. 
\end{align}
This result corresponds to the well known Casimir energy for a massless scalar field for conducting plates.

The latter limit is obtained in the weak coupling regime, where $\Lambda, \Lambda' \to 0$, and the mirrors introduce only a small perturbation to the free vacuum, i.e.
\begin{align}
\lim _{\Lambda, \Lambda' \ll 1} \mathcal{E} _{C} (L , \Lambda , \Lambda' ) & \approx  - \frac{ \hbar c }{ 24 \pi ^{2} L ^{3} } \Lambda \Lambda' \,  \int _{0} ^{\infty}  d \chi \, \chi \, e ^{-2 \chi }  = - \frac{ \hbar c }{ 96 \pi ^{2} L ^{3} } \Lambda \Lambda' . 
\end{align}
In order to verify that the divergent contribution to the Casimir energy has no physical effect, it is instructive to analyze the vacuum expectation value of the normal-normal component of the stress tensor, which determines the force per unit area acting on the plates. Unlike the energy density, the stress involves derivatives of the Green's function evaluated at the boundaries, and therefore provides a direct probe of the physically measurable pressure. By explicitly computing $\langle T ^{zz} \rangle$ we will show that the divergent term found in the energy density does not contribute to the force, leaving only the finite part to determine the Casimir pressure. From Eq.  (\ref{VEV_energy_momentum_reduced}) we obtain the corresponding VEV of the normal-normal component of the stress:
\begin{align}
    \langle T ^{zz} \rangle = \frac{1}{2i} \, \lim _{z \rightarrow z'} \, \int \frac{d \omega}{2 \pi } \int \frac{d ^{2} \mathbf{k} _{\perp} }{( 2 \pi ) ^{2} } \,  ( \gamma ^{2} + \partial _{z} \partial _{z'} )  \,  g \left( z , z' ; \omega , \mathbf{k} _{\perp} \right) . \label{VEV_Tzz}
\end{align}
Denoting by $\langle T _{zz} \rangle _{\parallel}$ the vacuum stress of the confined scalar field between the plates, and by $\langle T_{zz}\rangle _{\vert}$ the corresponding stress in the exterior region (due to only one of the plates), the net Casimir pressure follows from their difference:
\begin{equation}
\mathcal{P} _{C} (L) = \langle T_{zz} \rangle _{\parallel} \vert _{z=L } - \langle T_{zz} \rangle _{\vert} \vert _{z=L } . \label{CasPressure}
\end{equation}
As in the computation of the Casimir energy, the stress is obtained by substituting the appropriate reduced Green’s function into Eq.~(\ref{VEV_Tzz}). For the region between two confining plates, we use the reduced Green’s function in Eq.~(\ref{Reduced_Green_final}). Inserting this into Eq.~(\ref{VEV_Tzz}) and evaluating at the boundary $z=L$ yields
\begin{align}
    \langle T_{zz} \rangle _{\parallel} \vert _{z=L}  =   \frac{1}{2} \int \frac{d \zeta}{2 \pi } \int \frac{d ^{2} \mathbf{k} _{\perp} }{( 2 \pi ) ^{2} } \, \zeta \, \frac{ \lambda  \lambda ' +  (2 \zeta + \lambda ) ( 2 \zeta + \lambda ' ) \, e ^{2 \zeta  L}  }{ \lambda  \lambda ' -  (2 \zeta + \lambda ) ( 2 \zeta + \lambda ' ) \, e ^{2 \zeta  L}  } ,
\end{align}
where a Wick rotation $\omega\to i\zeta$ has been performed. To simplify the expression, we introduce polar coordinates in the $(\zeta,k_\perp)$-plane via $\xi =\sqrt{\zeta^{2}+k_\perp^{2}}$, with $\zeta = \xi \cos \varphi$, $k _{\perp} = \xi \sin \varphi$, integrate over the angle $\varphi$, and rescale to the dimensionless variable $\chi = L \xi$. Defining the dimensionless couplings $\Lambda=\lambda L$ and $\Lambda' = \lambda' L$, and restoring $\hbar$ and $c$, we obtain the compact representation
\begin{align}
    \langle T_{zz} \rangle _{\parallel} \vert _{z=L}  =     \frac{\hbar c}{4 \pi ^{2} L ^{4} } \int _{0} ^{\infty} d \chi \; \chi ^{3} \, \, \frac{ \Lambda  \Lambda ' +  (2 \chi + \Lambda ) ( 2 \chi + \Lambda ' ) \, e ^{2 \chi }  }{ \Lambda  \Lambda ' -  (2 \chi + \Lambda ) ( 2 \chi + \Lambda ' ) \, e ^{2 \chi }  } . 
\end{align}
We now consider the single-plate configuration. The corresponding reduced Green's function can be obtained directly from the two-plate result in Eq.~(\ref{Reduced_Green_final}) by switching off one of the $\delta$-interactions. Setting $\lambda = 0$ leaves a single semitransparent plate at $z=L$, and we obtain the
rank-one resolvent formula
\begin{align}
    g _{\vert} (z,z') = g _{0} (z,z') -  \frac{ g _{0} (z,L)  \, g _{0} (L,z') }{ \frac{1}{\lambda '} +  g _{0} (0,0) } .  \label{single_plate_Green}
\end{align}
Substituting Eq.~(\ref{single_plate_Green}) into Eq.~(\ref{VEV_Tzz}) and performing the same Wick rotation and angular integration as before yields the vacuum stress for the single-plate geometry:
\begin{align}
    \langle T_{zz} \rangle _{\vert} \vert _{z=L}  =  - \frac{1}{2} \int \frac{d \zeta}{2 \pi } \int \frac{d ^{2} \mathbf{k} _{\perp} }{( 2 \pi ) ^{2} } \, \zeta = - \frac{\hbar c}{4 \pi ^{2} L ^{4} } \int _{0} ^{\infty} d \chi \; \chi ^{3}  . 
\end{align}
Substituting the two-plate and single-plate
stresses derived above into Eq. (\ref{CasPressure}) we arrive at the compact representation
\begin{align}
\mathcal{P} _{C} (L) = \frac{\hbar c}{4 \pi ^{2} L ^{4} } \int _{0} ^{\infty} d \chi \; \chi ^{3} \, \left[ 1 -  \frac{    (2 \chi + \Lambda ) ( 2 \chi + \Lambda ' ) + \Lambda  \Lambda ' e ^{- 2 \chi }    }{   (2 \chi + \Lambda ) ( 2 \chi + \Lambda ' )   - \Lambda  \Lambda ' e ^{- 2 \chi }   }   \right] .   \label{Cas_Pressure}
\end{align}
It is straightforward to verify that the Casimir pressure can be obtained from the Casimir energy by differentiation with respect to the plate separation. Starting from Eq.~(\ref{finite_part}), the energy density depends on $L$ both explicitly, through the prefactor $L ^{-3}$, and implicitly, through the dimensionless couplings $\Lambda = \lambda L$ and $\Lambda '= \lambda 'L$. Taking the derivative with respect to $L$ at fixed $\lambda , \lambda '$ yields
\begin{align}
    \mathcal{P} _{C} (L) = - \frac{\partial \mathcal{E} _{C} (L) }{\partial L} . 
\end{align}
A short calculation shows that the derivatives of the couplings satisfy the identity
\begin{align}
    \Lambda \partial _{\Lambda} F (\chi ; \Lambda,\Lambda') + \Lambda ' \partial _{\Lambda '} F (\chi ; \Lambda,\Lambda') = -2 \, F (\chi ; \Lambda,\Lambda') 
\end{align}
where $F (\chi ; \Lambda,\Lambda') $ denotes the integrand in Eq.~(\ref{finite_part}). Substituting this result simplifies the derivative to exactly reproduce the integral expression for the pressure, Eq.~(\ref{CasPressure}). This confirms that the stress calculation and the energy derivative approach are fully consistent.

We now show directly from Eq.~(\ref{divergent_part}) that the divergent contribution to the energy does not generate any force. Taking the derivative of Eq.~(\ref{divergent_part}) with respect to $L$ at fixed $\lambda , \lambda'$ yields:
\begin{align}
- \frac{ \partial \mathcal{E}_{\infty} (L) }{ \partial L }  &= \frac{\hbar c}{6\pi ^{2}L ^{4}} \int _{0} ^{\infty} d \chi \; \frac{d}{d\chi} \left\{ \chi ^{3} (1 - e ^{- 2 \chi}) \, \ln \Big[(\Lambda+2\chi)(\Lambda'+2\chi)-\Lambda\Lambda' e^{-2\chi}\Big] \right\} .
\end{align}
The integrand is a total $\chi$-derivative, so only the endpoints contribute. At $\chi \to 0$, the prefactor behaves as $\chi ^{3} (1-e ^{-2\chi}) = \mathcal{O}(\chi ^{4})$, while the logarithm is finite, so the lower endpoint vanishes. At $\chi \to \infty$, the exponential term dies out, the logarithm grows at most as $\ln \chi ^{2}$, but is multiplied by $\chi^{3}(1-e^{-2\chi})$ inside a total derivative; the corresponding boundary term also vanishes. Therefore,
\begin{align}
    \mathcal{P}_{\infty}(L)
= - \frac{ \partial \mathcal{E}_{\infty} (L) }{ \partial L } = 0 . 
\end{align}
Hence, the divergent (surface) part of the energy is non-dynamical and does not contribute to the measurable Casimir pressure.

The results presented in this Section provide a complete and exact treatment of the Casimir effect in the presence of semitransparent plates, within a Lorentz-invariant framework. We obtained both the finite and divergent parts of the vacuum energy, including the surface contribution characteristic of $\delta$-like interactions, and analyzed their behavior in relevant physical limits. Furthermore, we explicitly verified that the divergent part does not contribute to the physical pressure: its derivative with respect to the plate separation vanishes, confirming that it corresponds only to surface self-energies. This conclusion was independently verified by evaluating the vacuum expectation value of the stress tensor on the plate, which reproduces the same result for the Casimir pressure.

In the next section, we extend this analysis to the case where Lorentz symmetry is explicitly broken, according to the modified field dynamics introduced in Section \ref{Model_Method_Section}.

\section{Casimir effect with semitransparent boundaries in a Lorentz-violating background} \label{Lorentz_violation_Section}

In order to investigate the Casimir effect in the presence of a Lorentz-violating background, the first step is to determine the corresponding reduced Green's function which satisfies
\begin{align}
  \left[  - h ^{0 0} \omega ^{2} - 2 h ^{0 i}   \omega k _{i} - h ^{i j}  k _{i} k _{j} - 2 i ( h ^{0 3}  \omega +  h ^{i 3} k _{i} ) \partial _{z} + h ^{3 3} \partial _{z} ^{2}  + \lambda \delta(z) + \lambda' \delta(z-L) \right] \mathfrak{g} \left( z , z' \right)  = \delta (z-z')  . 
\end{align}
The indices $i,j=1,2$ label directions parallel to the plates, i.e., transverse to the normal coordinate $z$. 

To construct the reduced Green's function in the presence of Lorentz symmetry violation, we follow the same strategy outlined in the previous Section. Specifically, we exploit the localized nature of the boundary interactions (modeled by Dirac delta functions) and express the full Green's function $\mathfrak{g} \left( z , z' \right)$ in terms of the free Green's function $\mathfrak{g} _{0} \left( z , z' \right)$, which now satisfies the modified differential equation
\begin{align}
  \left[ - h ^{0 0} \omega ^{2} - 2 h ^{0 i}   \omega k _{i} - h ^{i j}  k _{i} k _{j} - 2 i ( h ^{0 3}  \omega +  h ^{i 3} k _{i} ) \partial _{z} + h ^{3 3} \partial _{z} ^{2} \right] \mathfrak{g} _{0} \left( z , z' \right)  = \delta (z-z')  . \label{Reduced_Green_LV}
\end{align}
The solution is:
\begin{align}
    \mathfrak{g} _{0} \left( z , z' \right) = - \frac{i}{2 \xi h ^{33}} e ^{i \beta (z-z')}  e ^{i \xi (z _{>} - z _{<}) } , \label{Free_GF_Lorentz_Violation}
\end{align}
where $z_{<} = \mbox{min} (z,z')$, $z_{>} = \mbox{max} (z,z')$ and
\begin{align}
    \beta = \frac{1}{ h ^{33} } ( h ^{0 3}  \omega +  h ^{i 3} k _{i} ) , \qquad \gamma ^{2} = h ^{0 0} \omega ^{2} +  2 h ^{0 i}   \omega k _{i} + h ^{i j}  k _{i} k _{j} , \qquad \xi = \frac{1}{\vert  h ^{33} \vert }  \sqrt{ ( h ^{0 3}  \omega +  h ^{i 3} k _{i} )  ^{2} - h ^{33} \gamma ^{2} } . \label{definitions}
\end{align}
The complete solution for the reduced Green's function in the presence of the Dirac delta interactions follows the same structure as Eq. (\ref{Reduced_Green_final}), with the only modification being the replacement of the free Green's function $g _{0} \left( z , z' \right)$ by $\mathfrak{g} _{0} \left( z , z' \right)$. In other words, the general form of the solution remains unchanged, but it now incorporates the effects of Lorentz symmetry violation through the modified free propagator, i.e.

\begin{align}
    \langle T ^{00} \rangle &= - i  \lim _{z \rightarrow z'} \int \frac{d \omega}{2 \pi } \int \frac{d ^{2} \mathbf{k} _{\perp} }{( 2 \pi ) ^{2} } \, \left[  h ^{0 0} \omega ^{2} + h ^{0 i} \omega k _{i} + i \omega h ^{0 3}  \partial _{z} \right]  \mathfrak{g} \left( z , z' \right) - \langle \mathcal{L} \rangle , \label{VEV_energy_LV}
\end{align}
where the VEV of the Lagrangian can be expressed as
\begin{align}
    \langle \mathcal{L} \rangle = - \frac{i}{2} \lim _{z \rightarrow z'} \int \frac{d \omega}{2 \pi } \int \frac{d ^{2} \mathbf{k} _{\perp} }{( 2 \pi ) ^{2} } \, \left[ \gamma ^{2} + h ^{33} \partial _{z} \partial _{z ^{\prime}} + i h ^{33}  \beta   (\partial _{z ^{\prime} } - \partial _{z} ) \right] \mathfrak{g} \left( z , z' \right) . 
\end{align}
As expected, in the Lorentz-symmetric limit, the tensor (\ref{VEV_energy_LV}) reduces to its standard form (\ref{VEV_energy_LS}), thus recovering the known structure from the Lorentz-invariant theory. 

In the previous section, we deferred the proof that the VEV of the Lagrangian vanishes, reserving it for the present discussion. We now turn our attention to this point and provide an explicit demonstration. Using the reduced Green's function (\ref{Free_GF_Lorentz_Violation}) one finds
\begin{align}
    \lim _{z \to z'} \partial _{z} \partial _{z'} \mathfrak{g} _{0} \left( z , z' \right) = - \frac{i}{2 \xi h ^{33}} ( \beta +  \xi ) ^{2} ,  \qquad  \lim _{z \to z'}     ( \partial _{z} - \partial _{z'} ) \mathfrak{g} _{0} \left( z , z' \right) =  \frac{1}{  \xi h ^{33}}  \, ( \beta +  \xi )   ,
\end{align}
and therefore
\begin{align}
    \lbrace \mathcal{L} \rbrace _{v } =  \frac{i}{2} \lim _{z \rightarrow z'} \int \frac{d \omega}{2 \pi } \int \frac{d ^{2} \mathbf{k} _{\perp} }{( 2 \pi ) ^{2} } \,  \frac{i}{2 \xi h ^{33}} \left[ \gamma ^{2} + h ^{33} \left( \xi ^{2} - \beta ^{2}  \right)  \right] \mathfrak{g}_0 \left( z , z' \right) = 0 . 
\end{align}
The vanishing of this expression follows from the definitions of $\gamma$, $\beta$ and $\xi$. Here, the notation $\lbrace A \rbrace _{ v } $ denotes the vacuum expectation value of the operator $A$ evaluated in the absence of the plates, using the Green's function corresponding to the Lorentz-violating case in vacuum. In the Lorentz-symmetric limit $\lbrace A \rbrace _{v } \to \langle A \rangle _{v } $.

Consequently, the same conclusion, for $\langle A \rangle _{ v } $, holds in the Lorentz-symmetric case. Making use of this result, together with the relation $\lim _{z \to z'} \partial _{z} \mathfrak{g} _{0} \left( z , z' \right) =  i   ( \beta +  \xi ) \mathfrak{g} _{0} \left( z , z \right)$, the vacuum expectation value takes the form:
\begin{align}
    \lbrace T ^{00} \rbrace _{v } &= - i  \int \frac{d \omega}{2 \pi } \int \frac{d ^{2} \mathbf{k} _{\perp} }{( 2 \pi ) ^{2} } \, \left[  h ^{0 0} \omega ^{2} + h ^{0 i} \omega k _{i} - \omega h ^{0 3}   ( \beta +  \xi ) \right] \mathfrak{g} _{0} \left( z , z \right)  .
\end{align}
It should be noted that the last term, proportional to $h ^{03} \omega \xi$, vanishes upon integration due to symmetry considerations. The evaluation of this integral is hindered by the presence of a nontrivial quadratic form in $\xi = \frac{1}{\vert  h ^{33} \vert }  \sqrt{ ( h ^{0 3}  \omega +  h ^{i 3} k _{i} )  ^{2} - h ^{33} \gamma ^{2} }$, with $\gamma ^{2} = h ^{0 0} \omega ^{2} +  2 h ^{0 i}   \omega k _{i} + h ^{i j}  k _{i} k _{j}$. To overcome this difficulty, we introduce a change of variables that diagonalizes the quadratic form, thereby reducing the problem to a more tractable one. This transformation significantly simplifies both the analysis and the subsequent evaluation of the integral. The procedure is as follows. We first define the three-vector $\boldsymbol{\kappa} = (\omega , \mathbf{k} _{\perp})$, such that $\xi^{2} = \Lambda_{ij} \kappa^{i} \kappa^{j}$, where $\boldsymbol{\Lambda} = [\Lambda_{ij}]$ is a real, symmetric $3 \times 3$ matrix whose explicit form is directly determined from the definition of $\xi ^{2}$. At this point, we invoke Jacobi’s theorem, which ensures that any quadratic form in $n$ variables can be brought to diagonal form through an orthogonal transformation. Accordingly, we perform the change of variables $\kappa ^{\prime \, i} = \Psi _{ij} \kappa ^{j}$ with $\boldsymbol{\kappa}' = (\omega ', \mathbf{k}' _{\perp})$, where the orthogonal matrix $\boldsymbol{\Psi} = [\Psi _{ij}]$ is constructed from the normalized eigenvectors of $\boldsymbol{\Lambda}$. In the new primed coordinates the quadratic form becomes diagonal, i.e. $\xi ^{2} = \Pi _{ij} \kappa ^{\prime \, i} \kappa ^{\prime \, j} \equiv \Pi _{0} \omega ^{\prime \, 2} + \Pi _{1} k ^{\prime \, 2} _{x} + \Pi _{2} k ^{\prime \, 2} _{y} $, where $\Pi _{ij} = \Lambda_{in} \Psi _{nm} \Psi _{mj} \equiv \mbox{diag} (\Pi _{0},\Pi _{1},\Pi _{2})$. The intermediate steps of this calculation with the explicit coordinate transformation, carried out with Mathematica, lead to extremely lengthy expressions that are not particularly illuminating. For this reason, we refrain from explicitly displaying them and only report the final result. After performing a Wick rotation, we obtain
\begin{align}
    \lbrace T ^{00} \rbrace _{v } &= \frac{1}{\sqrt{ - h}} \int \frac{d \zeta ' }{2 \pi } \int \frac{d ^{2} \mathbf{k}' _{\perp} }{( 2 \pi ) ^{2} } \,  \frac{\zeta ^{\prime 2}}{ 2 \tilde{\gamma} '  }   , \label{VEV_T00_Lorentz_violation}
\end{align}
where $h = \det (h _{\mu \nu})$ and $\tilde{\gamma} ' = \sqrt{ \zeta ^{\prime 2} + k _{\perp} ^{\prime 2} } $. The integral appearing in this expression corresponds to the vacuum expectation value in the Lorentz-symmetric case (\ref{VEV_T00_Lorentz_symmetric}), and therefore the following identity holds:
\begin{align}
    \lbrace T ^{00} \rbrace _{v } &= \frac{1}{\sqrt{ - h}}  \, \langle T ^{00} \rangle _{v} . 
\end{align}
With the vacuum case fully determined, the next step is to compute the expectation value in the configuration where the plates are present, which will allow us to quantify the boundary contributions. In this case, we find
\begin{align}
    \lbrace T ^{00} \rbrace _{ \parallel } - \lbrace T ^{00} \rbrace _{v } &=  - i  \lim _{z \rightarrow z'} \int \frac{d \omega}{2 \pi } \int \frac{d ^{2} \mathbf{k} _{\perp} }{( 2 \pi ) ^{2} } \, \left[  h ^{0 0} \omega ^{2} + h ^{0 i} \omega k _{i} + i \omega h ^{0 3}  \partial _{z} \right]  \delta \mathfrak{g} \left( z , z' \right) ,  \label{VEV_T00_Lorentz_violation_plates}
\end{align}
where
\begin{align}
    \delta \mathfrak{g} \left( z , z' \right) = & \mathfrak{g} \left( z , z' \right) - \mathfrak{g} _{0} \left( z , z' \right) = - \frac{1}{\Delta} \lambda \, \{[1 + \lambda ' \, \mathfrak{g} _{0}(0,0)] \, \mathfrak{g} _{0}(z,0)  - \lambda ' \, \mathfrak{g} _{0} (0,L) \, \mathfrak{g} _{0}(z,L) \} \,  \mathfrak{g} _{0} (0,z') \notag \\[4pt] & \hspace{2.8cm} - \frac{1}{\Delta} \lambda ' \, \{ - \lambda \, \mathfrak{g} _{0} (0,L) \, \mathfrak{g} _{0} (z,0) + [ 1 + \lambda \, \mathfrak{g} _{0} (0,0) ] \, \mathfrak{g} _{0} (z,L) \} \, \mathfrak{g} _{0} (L,z') . \label{Delta_g}
\end{align}
Here, $\Delta = [ 1 + \lambda \, \mathfrak{g} _{0} (0,0)] \, [1 + \lambda ' \, \mathfrak{g} _{0} (0,0) ] - \lambda \, \lambda ' \, \mathfrak{g} _{0} (0,L) \, \mathfrak{g} _{0} (0,L)$. To evaluate the integral in Eq. (\ref{VEV_T00_Lorentz_violation_plates}) we use  the following results (for $0<z,z'<L$)
\begin{align}
    \lim _{z' \to z} \partial _{z} \, \mathfrak{g} _{0}(z,0) \,  \mathfrak{g} _{0} (0,z') &= i (\beta + \xi ) \,  \mathfrak{g} _{0}(z,0) \,  \mathfrak{g} _{0} (0,z) , \quad \qquad \lim _{z' \to z} \partial _{z} \, \mathfrak{g} _{0}(z,L) \,  \mathfrak{g} _{0} (0,z') = i (\beta - \xi ) \,  \mathfrak{g} _{0}(z,L) \,  \mathfrak{g} _{0} (0,z) . 
\end{align}
With the help of these results we obtain
\begin{align}
    \lbrace T ^{00} \rbrace _{ \parallel } - \lbrace T ^{00} \rbrace_{v } &=   i    \int \frac{d \omega}{2 \pi } \int \frac{d ^{2} \mathbf{k} _{\perp} }{( 2 \pi ) ^{2} } \,   \frac{1}{\Delta} \Bigg\{\ \lambda \, \{[1 + \lambda ' \, \mathfrak{g} _{0}(0,0)] \, \Xi _{+} \, \mathfrak{g} _{0}(z,0)  - \lambda ' \, \mathfrak{g} _{0} (0,L) \, \Xi _{-} \, \mathfrak{g} _{0}(z,L) \} \,  \mathfrak{g} _{0} (0,z) \notag \\[4pt]  & \hspace{3cm} +  \lambda ' \, \{ - \lambda \, \mathfrak{g} _{0} (0,L) \, \Xi _{+} \, \mathfrak{g} _{0} (z,0) + [ 1 + \lambda \, \mathfrak{g} _{0} (0,0) ] \, \Xi _{-} \, \mathfrak{g} _{0} (z,L) \} \, \mathfrak{g} _{0} (L,z) \Bigg\}\ , \label{Intermediate}
\end{align}
where we introduced the quadratic forms $\Xi _{\pm} = h ^{0 0} \omega ^{2} + h ^{0 i} \omega k _{i} - \omega h ^{0 3} (\beta \pm \xi )$. In this expression, we observe that, as in the vacuum case, the terms proportional to $\omega \xi$ vanish upon integration. This occurs because the product of the Green's functions turns out to be independent of $\beta$. Therefore, we can safely take $\Xi _{\pm} \to h ^{0 0} \omega ^{2} + h ^{0 i} \omega k _{i} - \omega h ^{0 3} \beta $ in Eq. (\ref{Intermediate}). This leaves us with
\begin{align}
    \lbrace T ^{00} \rbrace _{ \parallel } - \lbrace T ^{00} \rbrace _{v } &=  - i   \int \frac{d \omega}{2 \pi } \int \frac{d ^{2} \mathbf{k} _{\perp} }{( 2 \pi ) ^{2} } \, \left[ h ^{0 0} \omega ^{2} + h ^{0 i} \omega k _{i} - \omega h ^{0 3} \beta \right] \delta \mathfrak{g} \left( z , z \right).
\end{align}
We next apply the same steps to the integral as those used previously in the absence of plates. In short, we introduce a change of variable which diagonalizes $\xi ^{2}$, such that $\xi ^{2} = \Pi _{0} \omega ^{\prime \, 2} + \Pi _{1} k ^{\prime \, 2} _{x} + \Pi _{2} k ^{\prime \, 2} _{y} $. After further simplifications one realizes the identity
\begin{align}
    \lbrace T ^{00} \rbrace _{ \parallel } - \lbrace T ^{00} \rbrace _{v } &=  \frac{1}{\sqrt{-h}} \left[ \langle T ^{00} \rangle _{\parallel} - \langle T ^{00} \rangle _{v} \right] ,
\end{align}
which also implies a direct relation between the vacuum energies:
\begin{align}
    \mathrm{E} _{C} (L) = \int _{0} ^{L} \left[ \lbrace T ^{00} \rbrace _{ \parallel } - \lbrace T ^{00} \rbrace _{v } \right] \, dz = \sqrt{ \frac{ h ^{33} }{ h } } \, E _{C} (\tilde{L}) , \label{Casimir_energy_LV}
\end{align}
where $E _{C} (L)$ is the Casimir energy density in the Lorentz-symmetric (\ref{Casimir_energy}) case and $\tilde{L} = L / \sqrt{-h _{33}}$ is a rescaled length.  The final result shows that the Casimir energy density in the Lorentz-violating scenario (\ref{Casimir_energy_LV}) is proportional to its Lorentz-symmetric counterpart (\ref{Casimir_energy}), with the plate separation effectively rescaled and an overall multiplicative factor in front. It is worth emphasizing that this expression is exact: at no stage did we assume the Lorentz-violating coefficients to be small, in contrast to the more conventional perturbative treatments motivated by the empirical absence of observed Lorentz violation. One can further verify that in the Lorentz-symmetric limit the expression for the Casimir energy density correctly reduces to the standard result since $\tilde{L} \to L$ and $\mathrm{E} _{C} (L)$ tends to Eq. (\ref{EC0}) as $h _{\mu \nu} \to \eta _{\mu \nu}$.

To further ensure the consistency of our analysis, we have also computed the Casimir pressure directly from the vacuum stress acting on the plates. This provides an independent route to the physical observable, complementary to the evaluation obtained by differentiating the vacuum energy with respect to the plate separation. As expected, both methods yield identical results, thereby confirming the robustness of our treatment in the presence of Lorentz-violating terms. In the presence of Lorentz violation, the expression for the vacuum stress is given by
\begin{align}
    \langle T ^{33} \rangle &= - \frac{i}{2} \lim _{z \rightarrow z'} \int \frac{d \omega}{2 \pi } \int \frac{d ^{2} \mathbf{k} _{\perp} }{( 2 \pi ) ^{2} } \, \left[   \gamma ^{2} +   i h ^{3 3} \beta \,  ( \partial _{z} +  \partial _{z'} )  -   h ^{3 3} \, \partial _{z} \partial _{z'}   \right]  g \left( z , z' ; \omega , \mathbf{k} _{\perp} \right) . \label{Tzz_LV}
\end{align}
In close analogy with the procedure followed for the evaluation of the Casimir energy, the calculation of the stresses can also be carried out by inserting the appropriate reduced Green's functions into Eq. (\ref{Tzz_LV}) and performing once again the quadratic form diagonalization. The main distinction in this case lies in the fact that the Lagrangian density contributes explicitly to the stress tensor, introducing an additional layer of complexity compared to the energy computation. Nevertheless, by carefully applying the same methodology and making use of the Green function with the two mirrors $\mathfrak{g} \left( z , z' \right)$ given by Eq. (\ref{Delta_g}) and the corresponding expression of the Green's function with one mirror (which is obtained from $\mathfrak{g} \left( z , z' \right)$ by taking $\lambda = 0$) , one arrives after a direct, though somewhat lengthy, computation at the following expression:
\begin{align}
    \mathrm{P} _{C} (L) = \lbrace T ^{33} \rbrace _{ \parallel } \Big| _{z=L} - \lbrace T ^{33} \rbrace _{\vert} \Big| _{z=L} = - \frac{\partial \mathrm{E} _{C} (L) }{\partial L} =  \frac{1 }{ \sqrt{-h } } \, \mathcal{P} _{C} (\tilde{L}) , 
\end{align}
where $\mathcal{P} _{C} (L)$ is the Lorentz-symmetric Casimir pressure, given by Eq. (\ref{Cas_Pressure}).

In the following section, we discuss possible physical realizations of the model and present illustrative plots of the Casimir energy using parameter values motivated by these scenarios.

\section{Physical realizations and numerical illustrations} \label{Physical_realizations}

In this section, we discuss possible physical realizations of the theory, which allow us to estimate realistic values for the model parameters. In particular, we focus on two sets of parameters: the transparency of the mirrors, characterized by $\lambda$ and $\lambda^\prime$, and the components of the background tensor $h_{\mu\nu}$ that parametrizes Lorentz symmetry violation. For the mirrors, we consider thin dielectric slabs of various materials and thicknesses, which provide concrete estimates for $\lambda$ and $\lambda^\prime$ based on their refractive indices and thicknesses. Regarding the Lorentz-violating tensor $h_{\mu\nu}$, we envision a phenomenological scenario inspired by analogue optical media, in which an anisotropic, birefringent material effectively simulates the presence of a preferred background. This setup allows us to assign plausible magnitudes to the components of $h_{\mu\nu}$, providing a physically motivated range for our analysis. These parameter estimates will then be used to generate illustrative plots of the Casimir energy, exploring both the influence of mirror transparency and Lorentz-violating effects.

We first discuss the parameters $\lambda$ and $\lambda '$. Consider a dielectric slab of thickness $d$ and refractive index $n$ placed in vacuum. In the limit where $d$ is small compared to the relevant wavelengths, the interaction of a scalar field with the slab can be modeled by a delta-function potential, with strength
\begin{align}
    \lambda \simeq \frac{n ^{2}-1}{d} , \qquad \lambda ' \simeq \frac{n ^{\prime 2}-1}{d '} ,
\end{align}
where $n$ and $n '$ are the refractive indices of the two slabs and $d, d'$ their thicknesses. These expressions arise from integrating the standard wave equation across the thin slab and matching the discontinuity in the derivative of the field. Physically, this corresponds to approximating the slab as an infinitesimally thin mirror whose reflectivity is proportional to the contrast in dielectric permittivity, $n ^{2} - 1$.

As a concrete numerical estimate, we consider silica ($\mathrm{SiO_2}$) as the slab material, which has a refractive index $n \simeq 1.45$ in the visible spectrum \cite{palik2012handbook}. Choosing a thickness $d = 10\mathrm{nm}$, the corresponding coupling parameter becomes
\begin{align}
    \lambda \simeq \frac{(1.45) ^{2}-1}{10 \times 10 ^{-9} \mbox{m}} \simeq 1.10 \times 10 ^{8} \mbox{m} ^{-1} . 
\end{align}
For the second slab, we can take the same material and thickness, yielding $\lambda ' \simeq \lambda$. These values lie in the intermediate regime between fully transparent mirrors ($\lambda, \lambda ' \to 0$) and perfectly reflecting boundaries ($\lambda, \lambda ' \to \infty$), making them suitable for exploring semitransparent Casimir setups. Finally, to approach the nearly perfect mirror limit, one can consider a higher-index material such as silicon at optical frequencies ($n \simeq 3.5$ \cite{palik2012handbook}) with thickness $d = 10 \mathrm{nm}$: $\lambda \simeq 1.12 \times 10 ^{9} \mbox{m} ^{-1}$. Such estimates provide a physically meaningful scale for $\lambda$ and $\lambda '$, allowing us to generate illustrative plots of the Casimir energy as a function of the mirror separation and transparency. 

In order to provide a physically realistic scenario for the Lorentz-violating tensor $h _{\mu \nu}$, we consider the propagation of a scalar field in an anisotropic dielectric medium, specifically a uniaxial material with its optical axis aligned along the $z$-direction. In such a medium, the refractive index differs between the transverse plane ($x$-$y$) and the axis direction ($z$), denoted by $n _{\perp}$ and $n _{\parallel}$, respectively. This anisotropy naturally induces a preferred direction in the effective spacetime seen by the field, providing a concrete realization of Lorentz symmetry violation in a controlled laboratory setting. 

Within the Gordon metric approximation \cite{https://doi.org/10.1002/andp.19233772202, Matt_Visser_1998}, the propagation of the scalar field in this medium can be expressed in terms of an effective metric
\begin{align}
    g _{\mu \nu} ^{\mbox{\scriptsize eff}} = \mbox{diag} \left( n _{\perp} ^{2},-1,-1, \frac{n _{\perp} ^{2}}{n _{\parallel} ^{2}} \right) , 
\end{align}
which directly leads to the Lorentz-violating tensor components
\begin{align}
    h _{00} = n _{\perp} ^{2}, \quad h _{xx} = h _{yy} = - 1, \quad h _{zz} = - \frac{n _{\perp} ^{2}}{n _{\parallel} ^{2}} .
\end{align}
This construction ensures that wave propagation along the optical axis ($z$) differs from propagation in the transverse plane, realizing a nontrivial anisotropic background.

As a concrete example, we consider 5CB (pentyl-cyanobiphenyl), a commonly used nematic liquid crystal \cite{10.1063/1.1877815, de1993physics}. Typical refractive indices for 5CB at room temperature are $n _{\perp} \approx 1.50$ and $n _{\parallel} \approx 1.70$. In this case, the corresponding Lorentz-violating components read 
\begin{align}
    h _{00} = 2.25, \quad h _{xx} = h _{yy} = - 1, \quad h _{zz} = - 0.78 .
\end{align}
In this scenario, the liquid crystal provides a laboratory realization in which the scalar field experiences different effective propagation speeds along the optical axis and the transverse directions, giving concrete numerical estimates for $h ^{\mu \nu}$. These values can be used to generate illustrative plots of the Casimir energy between semitransparent mirrors and to study the influence of Lorentz-violating anisotropy on the vacuum energy in a controlled experimental setup.

\begin{figure}
    \centering
    \includegraphics[width=0.5\linewidth]{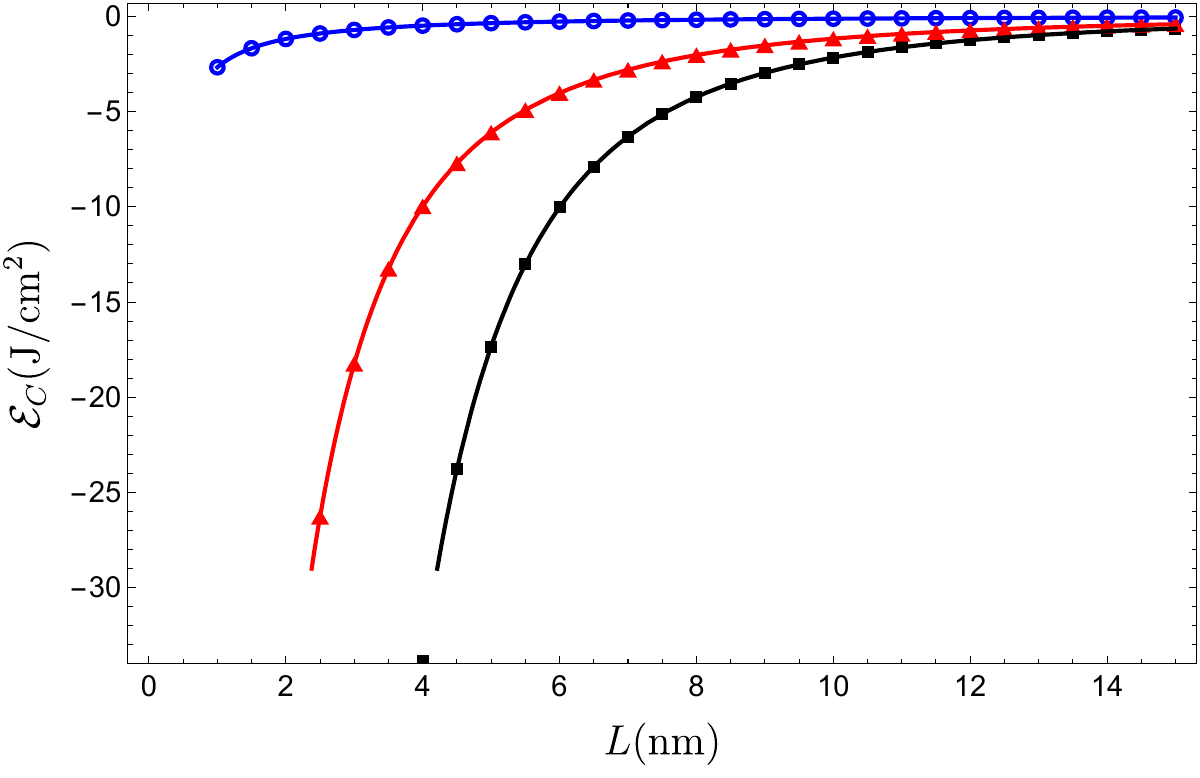} \,\,\,\,\,\,\,\,\,\,\,\,\,\, \includegraphics[width=0.44\linewidth]{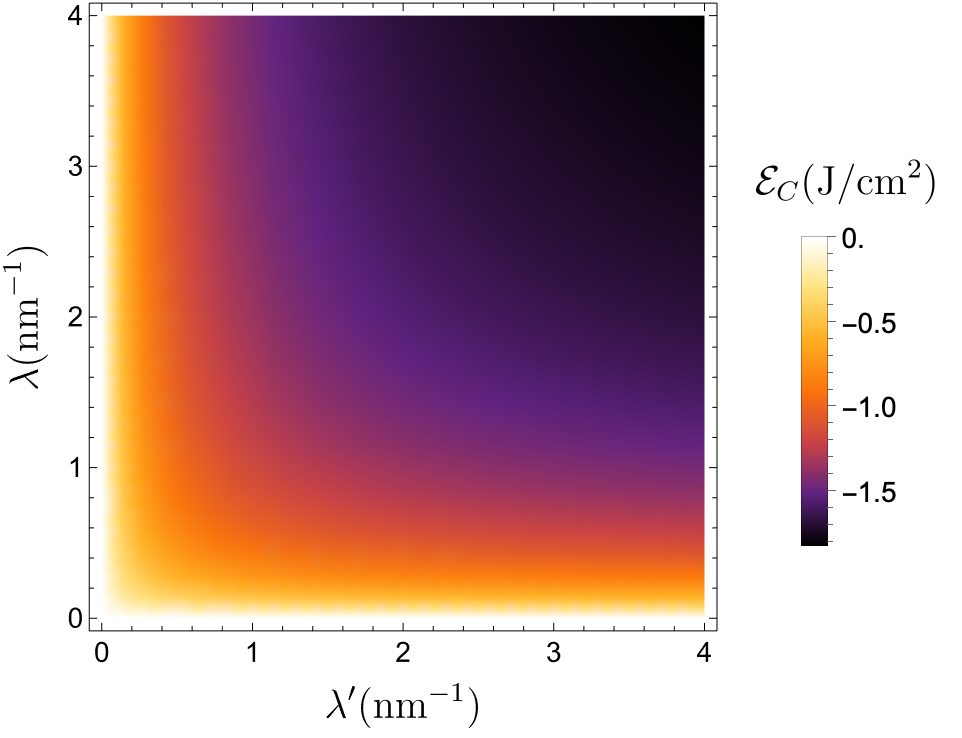}
    \caption{Left: Casimir energy density (in units of energy per unit area) as a function of plate separation (in nanometers). The black squares show the standard Lorentz-invariant result for perfectly reflecting boundaries. The red triangles indicate the case of nearly perfect mirrors, while the blue circles correspond to a more transparent regime, where the reduced reflectivity weakens the confinement of vacuum fluctuations and thus lowers the Casimir interaction. Right: Density plot of the Casimir energy as a function of the transparency parameters $\lambda$ and $\lambda '$, for a fixed plate separation. The upper-right region corresponds to the limit of perfectly reflecting mirrors, where the energy approaches the standard Casimir result. Along the diagonal ($\lambda = \lambda '$), the mirrors are equally transparent, while deviations from the diagonal highlight asymmetric configurations with different transparency strengths. The energy vanishes smoothly in the transparent limit ($\lambda , \lambda ' \to 0$). }
    \label{Cas_energy}
\end{figure}

In Fig. \ref{Cas_energy}, at left, we display the Casimir energy density (per unit area) as a function of the plate separation, measured in nanometers, for a system embedded in the nematic liquid crystal 5CB (pentyl-cyanobiphenyl). The vertical axis is expressed in units of energy per square centimeter. The black squares correspond to the standard Lorentz-invariant result with perfectly reflecting mirrors, which we use as a benchmark. The red triangles illustrate the nearly perfect mirror limit (silicon), showing only a mild suppression relative to the ideal case. The blue circles correspond to a more transparent configuration (silica), where the Casimir energy density is substantially reduced.

This trend reflects the role of boundary transparency in the confinement of vacuum modes. Perfectly reflecting plates enforce strict boundary conditions, producing the maximal shift in vacuum energy. As transparency increases, a larger fraction of modes is allowed to leak through the boundaries, thereby diminishing the modification of the vacuum state and reducing the strength of the Casimir interaction.

In Fig. \ref{Cas_energy}, on the right, we present the density plot of the Casimir energy as a function of the coupling parameters $\lambda$ and $\lambda '$, which characterize the transparency of the two mirrors. This representation makes it possible to analyze the full parameter space and identify the different physical regimes of the system. In the upper-right region of the plot, corresponding to large values of both couplings, the mirrors effectively behave as perfect reflectors and the Casimir energy approaches the standard result for two ideal plates. Along the diagonal, where $\lambda = \lambda '$, the mirrors are equally transparent and the energy decreases monotonically with decreasing coupling strength, interpolating between the ideal limit and the transparent regime.

Away from the diagonal, the asymmetric case $\lambda \neq \lambda '$ reveals a richer structure. When one mirror is nearly perfect ($\lambda \to \infty$) while the other is semi-transparent, the energy is suppressed but remains finite, reflecting the fact that both boundaries are required to confine the vacuum modes effectively. As both couplings decrease, the energy continuously vanishes, reaching the transparent limit ($\lambda , \lambda ' \to 0$) where no Casimir interaction survives.

Overall, this density plot highlights how the Casimir energy depends simultaneously on both transparency parameters. It illustrates the smooth interpolation between the ideal case and the transparent regime, while also showing the nontrivial asymmetry that emerges when the mirrors possess different optical properties. Such a global view makes it clear that the Casimir effect can be finely tuned through the independent adjustment of the couplings $\lambda$ and $\lambda '$.

\section{Conclusions} \label{Conclusions}

In this work we have analyzed the Casimir effect for a massless scalar field in the presence of Lorentz-violating modifications of the vacuum, confined between two parallel semitransparent mirrors. Starting from the scalar sector of the Standard-Model Extension (SME), we have shown that the combined effects of mirror transparency and Lorentz violation can be computed in closed analytical form using Green's function techniques. A key outcome of our analysis is that the two types of corrections factorize: the transparency of the mirrors and the Lorentz-violating background contribute multiplicatively to the Casimir energy and effectively modify the plate separation.

From the perspective of high-energy physics, Lorentz symmetry violation is expected to be extremely small, given the stringent experimental limits that exist at laboratory and astrophysical scales. However, when reinterpreted in the framework of condensed matter systems, the same SME-inspired parametrizations can acquire direct physical meaning. In anisotropic dielectrics and liquid crystals, for example, the refractive indices naturally encode preferred spatial directions, effectively generating nontrivial coefficients that mimic Lorentz-violating backgrounds. This dual interpretation is conceptually important: while the SME provides a theoretical extension of the Standard Model at fundamental scales, condensed matter realizations offer experimentally accessible platforms where the same structures can be probed and exploited \cite{PhysRevD.109.065005, sym17040581}.

The Casimir effect itself has historically played a central role in advancing both theoretical and experimental physics. Beyond its original role as a striking manifestation of vacuum fluctuations, it has become a precision tool for testing fundamental interactions, probing modifications of electrodynamics, and exploring the interplay between geometry, material properties and quantum fields. In recent decades, experimental advances have enabled accurate measurements of Casimir forces in diverse settings, including metallic and dielectric surfaces, semitransparent membranes, and systems with tunable boundary conditions. Our results extend this line of investigation by showing how the Casimir energy responds not only to imperfect boundaries but also to modifications of the underlying vacuum structure.

To illustrate the physical relevance of our approach, we provided estimates based on realistic material parameters. In particular, for the nematic liquid crystal 5CB (pentyl-cyanobiphenyl), which is widely used in liquid crystal displays and precision optics, we extracted effective values for the Lorentz-violating tensor components. These estimates show that, unlike in the high-energy context, the anisotropies in such condensed matter systems are not negligible but directly determined by the refractive indices of the medium. As a result, the modifications to the Casimir energy predicted in our framework could be significant in experimentally realizable setups involving parallel plates immersed in liquid crystals.

Taken together, our findings reinforce the idea that Lorentz-violating effective field theories are not only of interest in the search for new physics beyond the Standard Model, but also provide a versatile language to describe anisotropic and dispersive media in condensed matter. The Casimir effect, with its sensitivity to both boundary conditions and background modifications, serves as a natural arena to bridge these two perspectives. Future work may include extending the present analysis to fermion fields, exploring temperature effects, or designing concrete experimental proposals based on liquid crystals such as 5CB to test the predictions derived here.

\acknowledgements{A.M.-R. acknowledges financial support by DGAPA-UNAM Project No. IG100224, by SECIHTI project No. CBF-2025-I-1862 and by the Marcos Moshinsky Foundation.}

\bibliography{references.bib}

\end{document}